\documentclass[a4paper]{jpconf}
\usepackage{graphicx}
\usepackage{amsmath}
\usepackage{amssymb}

\begin{document}
\title{
Foreseeing Neutrino  spectra  in Deep Core}

\author{Daniele Fargion and  Daniele D'Armiento,\\ Paola Di Giacomo, Paolo Paggi}

\address{Physics Department, Rome University 1, INFN 1 Roma, Ple. A. Moro 2, Italy}

\ead{daniele.fargion@roma1.infn.it}

\begin{abstract}
Atmospheric muon neutrino in Deep Core (whose rate and spectra might be soon available) should exhibit an $''$anomaly$''$  (due to  tens GeV up-going muon neutrino suppression, converted into tau flavor) that must be imprinted in out-coming  rate spectra. We estimate here our independent  muon neutrino spectra based on SK and its projected record on Deep Core  Channels. Our estimate (based on cosmic rays, muon records and tested Super-Kamiokande (SK) data)  differs both in shape and in rate from other previous published spectra. The expected rate might exhibit a minimum near channel 6 of Deep Core strings and it should manifest strong  signature for flavor mixing (mostly between channel 4--15) and a relevant $''$anomaly$''$ for eventual CPT violation (MINOS like) written at channel 3--6, whose statistical weight (mainly at channel 5) might soon confirm or dismiss MINOS CPT claim \cite{1}. At the flux minimum  around channel 6, (a flux suppressed respect the non oscillated case at least by an order of magnitude) the atmospheric neutrino paucity offers a better windows to a twenty GeV Neutrino Astronomy. Therefore by doubling the string array we may foresee a richer rate and a more complete (zenith and azimuth) atmospheric neutrino distribution and an exciting first twenty GeV Astronomy.
\end{abstract}

\section{Introduction}
The role of neutrino muon and tau mixing among flavors has been observed since nearly 13 years and it has been foreseen already in different recent high energy astrophysical roles \cite{0},\cite{8},\cite{10}.
Recently  MINOS results hint for a CPT violation in neutrino sector. We considered therefore  a  long base experiment to Deep Core to disentangle such a neutrino CPT puzzle. The suggested  beaming (OPERA like experiment) of neutrinos from CERN (or FNL) across the Earth to  IceCube might provide a novel signature for muon and anti-muon neutrino mixing while better focusing flavor parameters \cite{6} . Nevertheless the present atmospheric neutrino noise that should arise in Deep Core data is also bearing important records on neutrino mixing and eventual CPT violation. Indeed we used previous (Deep Core)  expected rate to foresee (over the same expected spectra) deviation due to CPT (MINOS like) violations.  Nevertheless a more accurate study did lead us to foresee a different spectra \cite{7} , to be elaborated here, more in detail,  based on the known Super-Kamiokande detection, scaled to Deep Core volumes (and environment). Earlier \cite{7} preliminary result is here more deepened and confirmed, starting from primordial atmospheric spectra, source of atmospheric neutrinos up to their interactions and  muon  event rate whose tracks are projected along each channel string. Results are somehow surprising respect earlier ones.

\section{From the atmospheric cosmic ray to the event rate in Deep Core}

Let us remind the neutrino energy flux defined as
\begin{equation}
\phi_{\nu_{\mu}}(E_{\nu_{\mu}})\,=\,E_{\nu_{\mu}}^2\,\cdot\,\frac{dN_{\nu_{\mu}}}{dE_{\nu_{\mu}}\,dt\,dA\,d\Omega}
\end{equation}

that we parameterized, here for the first time, fitting the well known curve from refs.\cite{1,2} for muon neutrinos  with the following analytical formula:

\begin{equation}
\phi_{\nu}(E_{\nu})\,=\,10^2\,\left(\frac{1}{9.5\,E_{\nu}^{-1.12}}\,+\,\frac{1}{4}\,+\,\frac{1}{27\,E_{\nu}}\right)^{-1}\;\textrm{m}^{-2}\,\textrm{s}^{-1}\,\textrm{sr}^{-1}\,\textrm{GeV},
\end{equation}

where the energy $E_{\nu}$ is expressed in GeV unity; the energy range of this formula embrace the hundred MeV up to TeV. This averaged isotropic spectra must be  enhanced (empirically) at horizons (slightly at low, GeVs energy and more at high TeV energy)  because of  longer pion, kaons and muons decay  flight. We disregarded for Deep Core and IceCube at South Pole the  geomagnetic cut off influence; (its  role at lowest energy has been taken into account  at SuperKamiokande for azimuthal directions). To take into account of the enhancement (of horizontal respect to vertical rate)  we derived (to fit the SK data) a correction factor to be multiplied to previous flux:

\begin{equation}
f(E_{\nu},\theta)\,=\,\left[1\,+\,\left(\frac{E_{\nu}}{50}\right)^{\frac{1}{5}}\,(\sin\theta)^{0.7}\right]^{-1},
\end{equation}

where the energy is in GeV unity, and horizontal angle $\theta$ (complemental to zenith angle $\hat{z}$) range from $0$ (horizon) to $\pi /2$ (upward vertical direction).
The total mass target in Deep Core for incoming neutrino is taken from the effective volume estimated (for each energy) in \cite{3}, without online veto. The ratio between all (fully contained (FC) with partially contained (PC) and upward stopping and through going, i.e. without  on line veto) events versus only (FC+PC, i.e. on-line veto) is unity up to $\sim\,100$ GeV, reaching a ratio $\sim\,1.5$ at $400$ GeV, or $\sim\,2$ at TeV. In these highest energies the channel rate is low and it is not so relevant  in our neutrino oscillation study, mostly focused below $200$ GeV and below maximal 50 channels. Therefore we considered here all the events without veto, almost coincident with the ones with veto  within 50 channels count, within the following Deep Core energy dependent (as expected following \cite{3} in GeV energy unity), threshold volume:

\begin{equation}
V(E_{\nu})\,=\,1.17\cdot10^6\cdot(5.5\,E_{\nu}^{0.2}\,-\,5.5)\cdot\left[3.2\,+\,0.8\,\cdot\ln\left(\frac{E_{\nu}}{80}\right)\right]\;\textrm{m}^3.
\label{V}
\end{equation}

Neutrino Charge Current (CC) cross section for muon production is considered (see ref. \cite{5}) to be linear increasing with energy, in the energy range considered, while antineutrino cross section is taken about a half of the neutrino one. Electron neutrino and tau neutrinos CC and all flavor neutral current (NC) event are source of spherical showers whose role may influence (and pollute) only a few channel (3-5) and are only mentioned here: they will be  discussed in detail elsewhere.

The calculation includes neutrino oscillation probability as a function both of energy and arrival observation angle, given the earth chord distance for each angle below the horizon, $ P_{\nu_{\mu} \rightarrow \nu_{i}}(E_{\nu},\theta), i=1,2,3 $; it means that we take into account three exact neutrino flavor oscillation (see \cite{6}), discussing the out-coming (observable) muon tracks.  We took into account also the matter influence on the mixing inside the Earth assuming at each zenith angle the corresponding  average Earth density.  We will show elsewhere that the exact integral in variable Earth density do not change our present results at $10$--$100$ GeV ranges, but mostly induce changes at lower energies.

\begin{figure}[h]
\begin{minipage}{14pc}
\includegraphics[width=14pc]{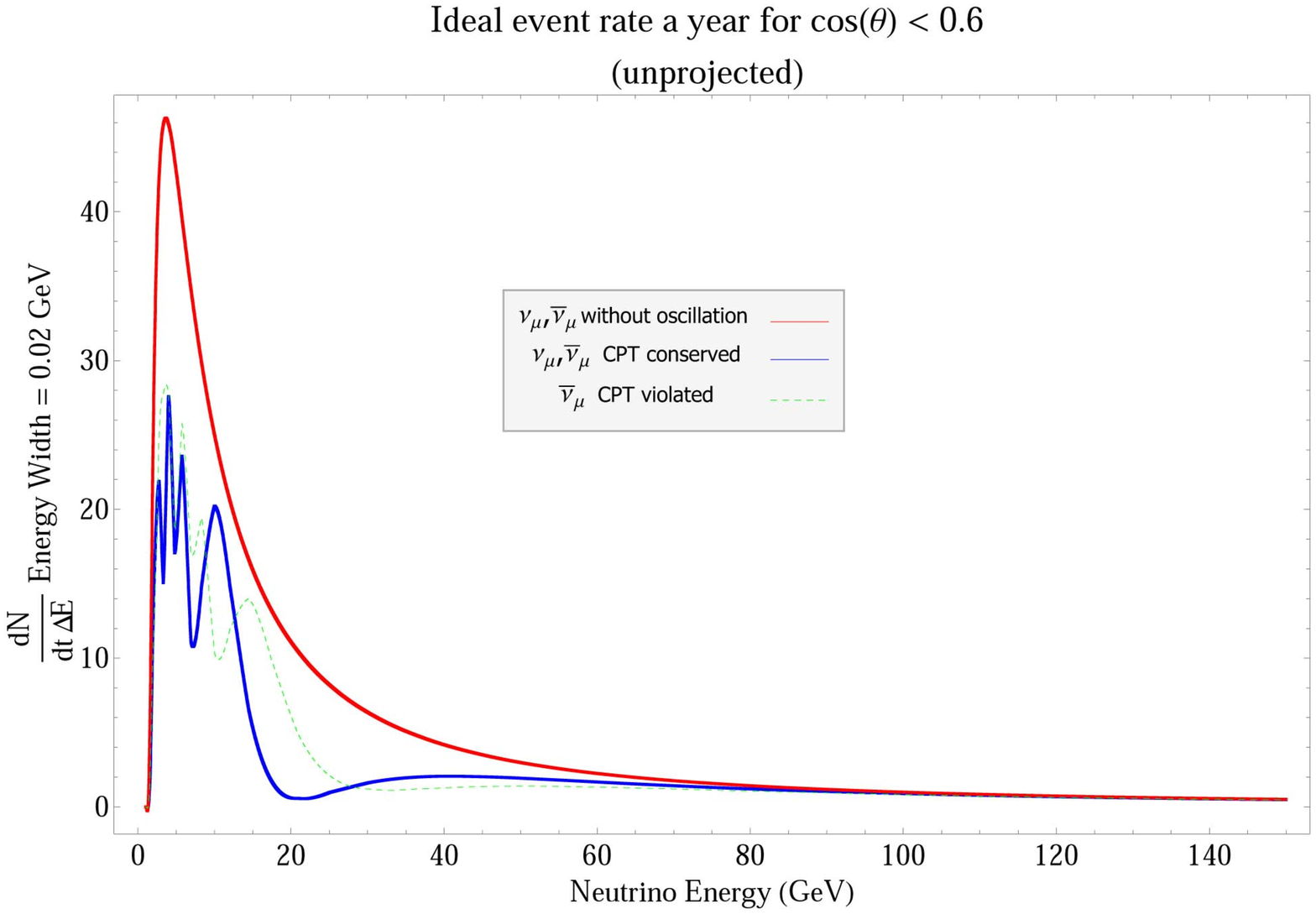}
\caption{\label{002GEV}The ideal count rate that Deep Core would be able to reveal  on the muon track length if it could
  disentangle muons within 20 MeV energy width.
It is clear the marked difference between non oscillated case (red non oscillating curve), the mixed CPT conserved case ( oscillating blue curve),
and the thin-dashed CPT violated (green) alternative signal. This is in vacuum case. }
\end{minipage}\hspace{2pc}%
\begin{minipage}{14pc}
\includegraphics[width=14pc]{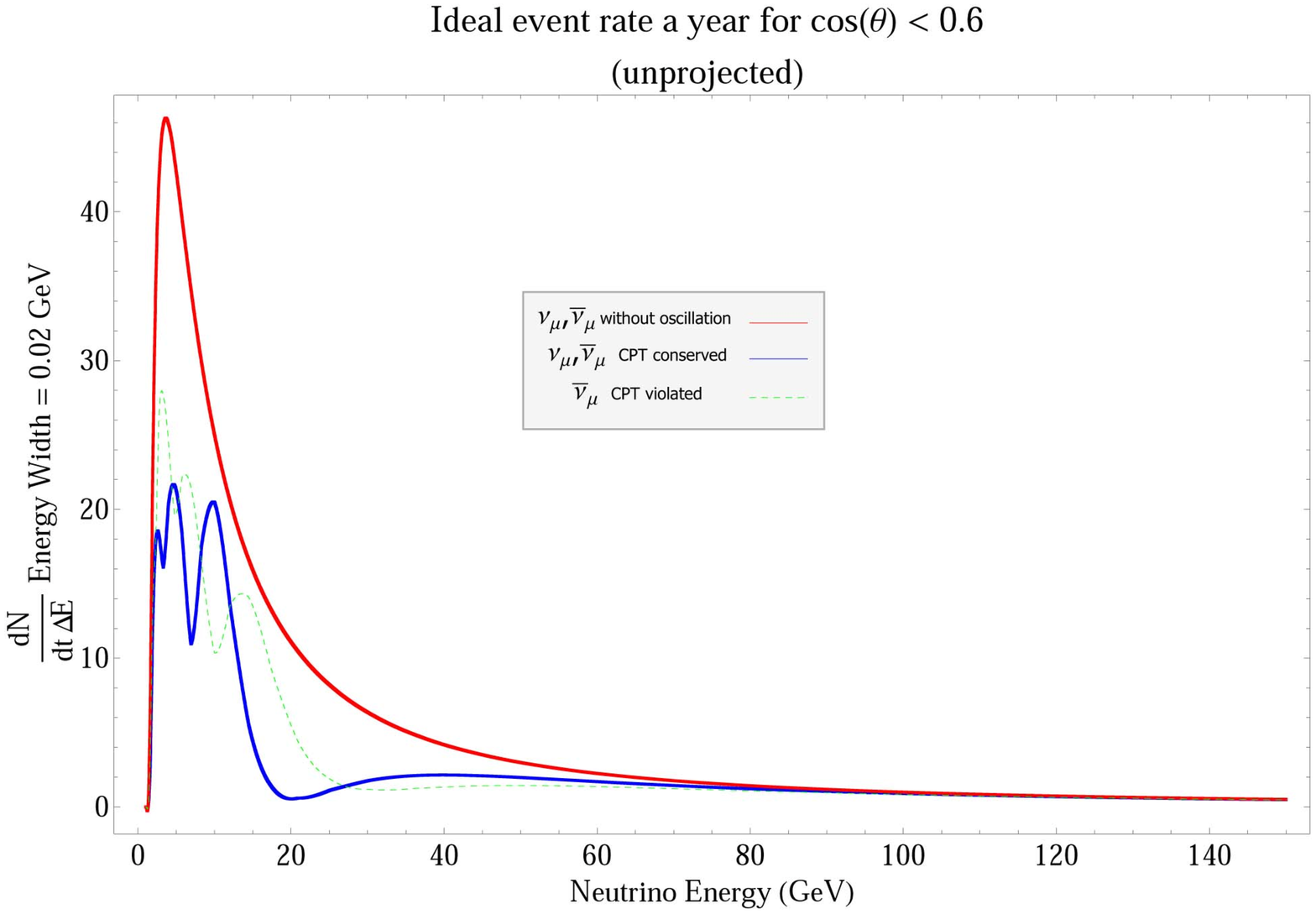}
\caption{\label{1GEV} As previous figure, where neutrino mixing occurs in matter. Note the little difference in spectra at the very low energies.}
\end{minipage}
\end{figure}

\begin{figure}[h]
\begin{minipage}{17pc}
\includegraphics[width=14pc]{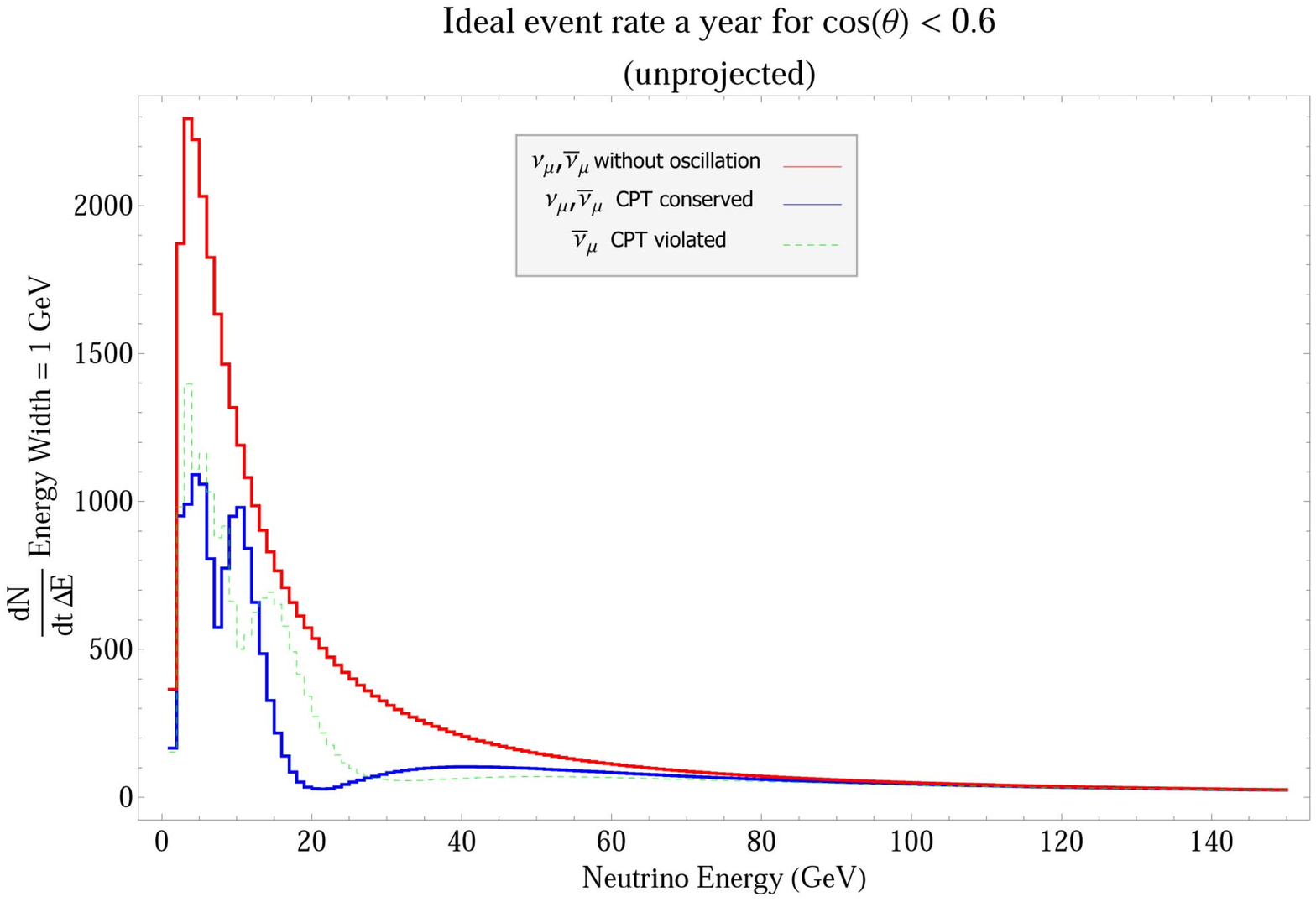}
\caption{\label{1GEV} As previous figure the averaged (within 1 GeV) spectra of atmospheric neutrino
if Deep Core would be able to disentangle within $1$ GeV energy width. As in previous curves, the non oscillated one is red and non oscillating , the mixed CPT conserved case is still oscillating and it is a blue curve, while the thin-dashed CPT violated (green) curve is an alternative signal. Most, but not all, of the low energy oscillations are lost.}
\end{minipage}\hspace{2pc}%
\begin{minipage}{17pc}
\includegraphics[width=14pc]{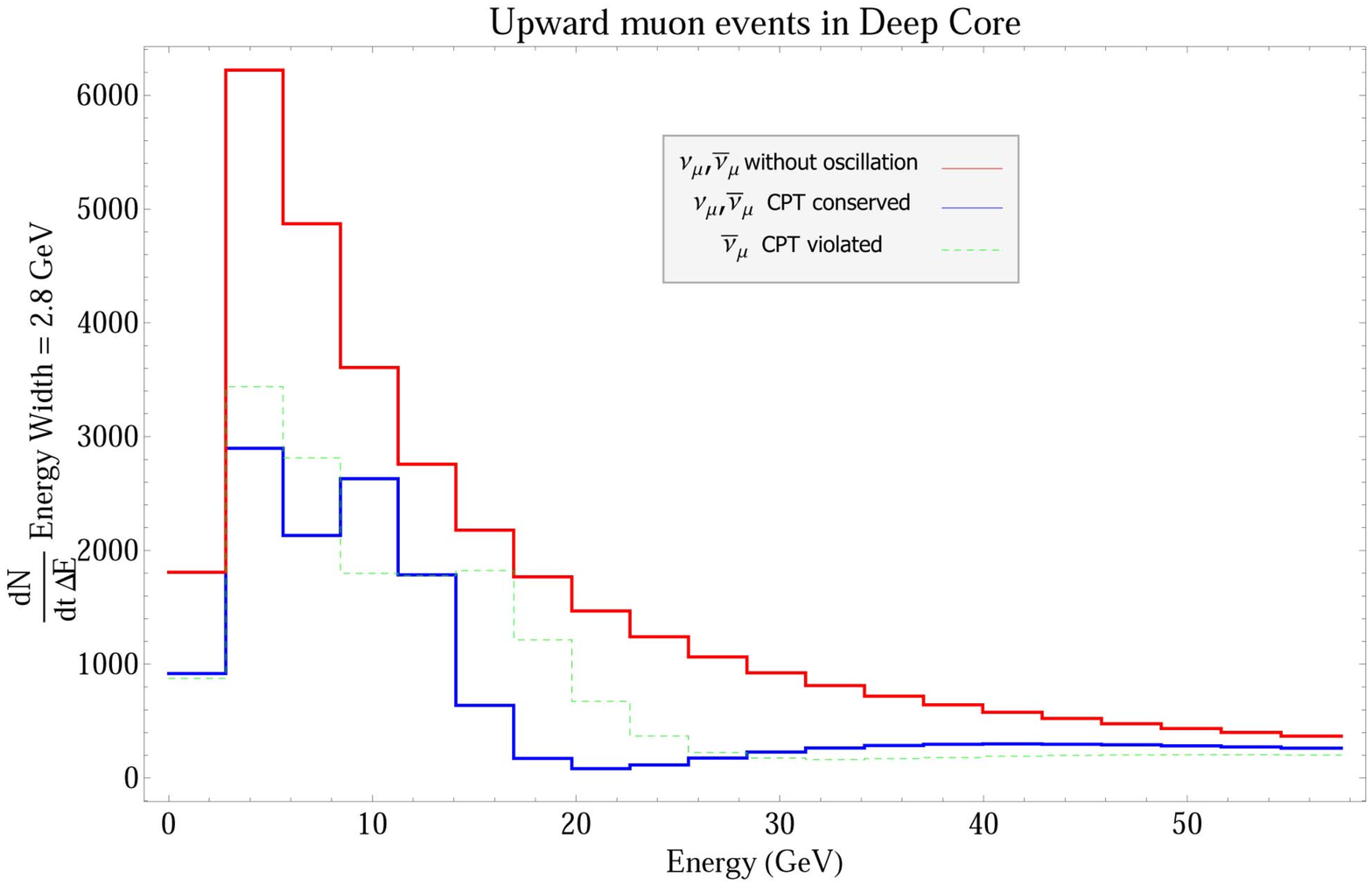}
\caption{\label{2e8GEV} A possible real count rate as a function of the energy within a $2.8$ GeV width (as each channel from a signal of vertical muon in Deep Core). As above the red curve marks the un-oscillated case, the blue one the CPT conserved version and the thin-dashed green line stands for the CPT violated solution. The smearing of the earliest oscillation is complete. Few mixing structure remains at low channels.}
\end{minipage}
\end{figure}

\begin{figure}[h]
\begin{center}
\includegraphics[width=22pc]{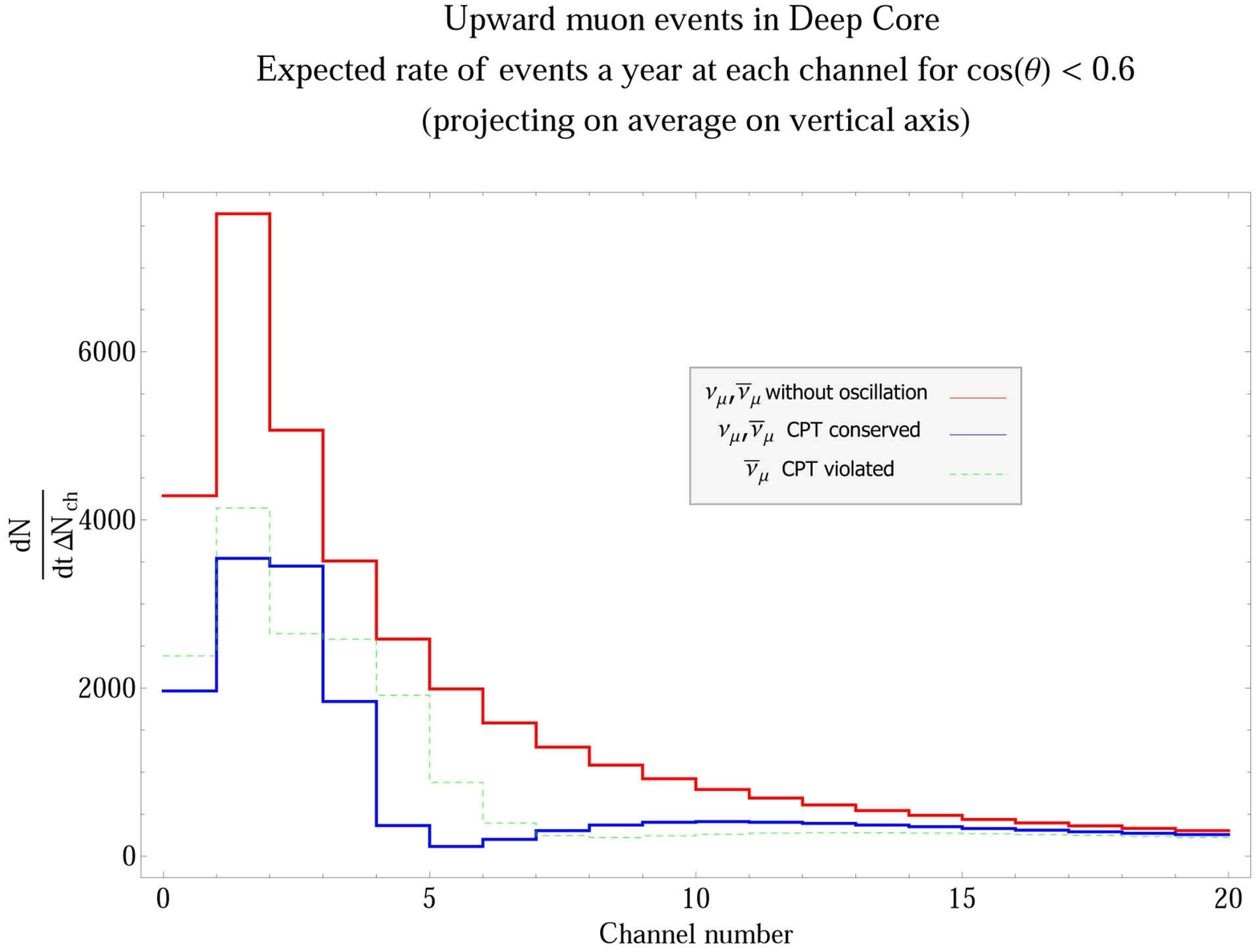}
\caption{\label{projected}The most probable realistic rate count rate a year as a function of the channel within a $3.88$ GeV width projected within a factor 0.72 (as the  channel distance in Deep Core). As above the red curve marks the un-oscillated case, the blue one the CPT conserved version and the thin-dashed green line stand for the CPT violated solution. The smearing of the earliest oscillation is hidden. But the deepening  structure remains at 6-9 channels remind of the muon disappearance at twenty GeVs. The depression of the muon flux in this region is nearly a factor ten, making this window silent and ideal to search for a Neutrino Astronomy}
\includegraphics[width=20pc]{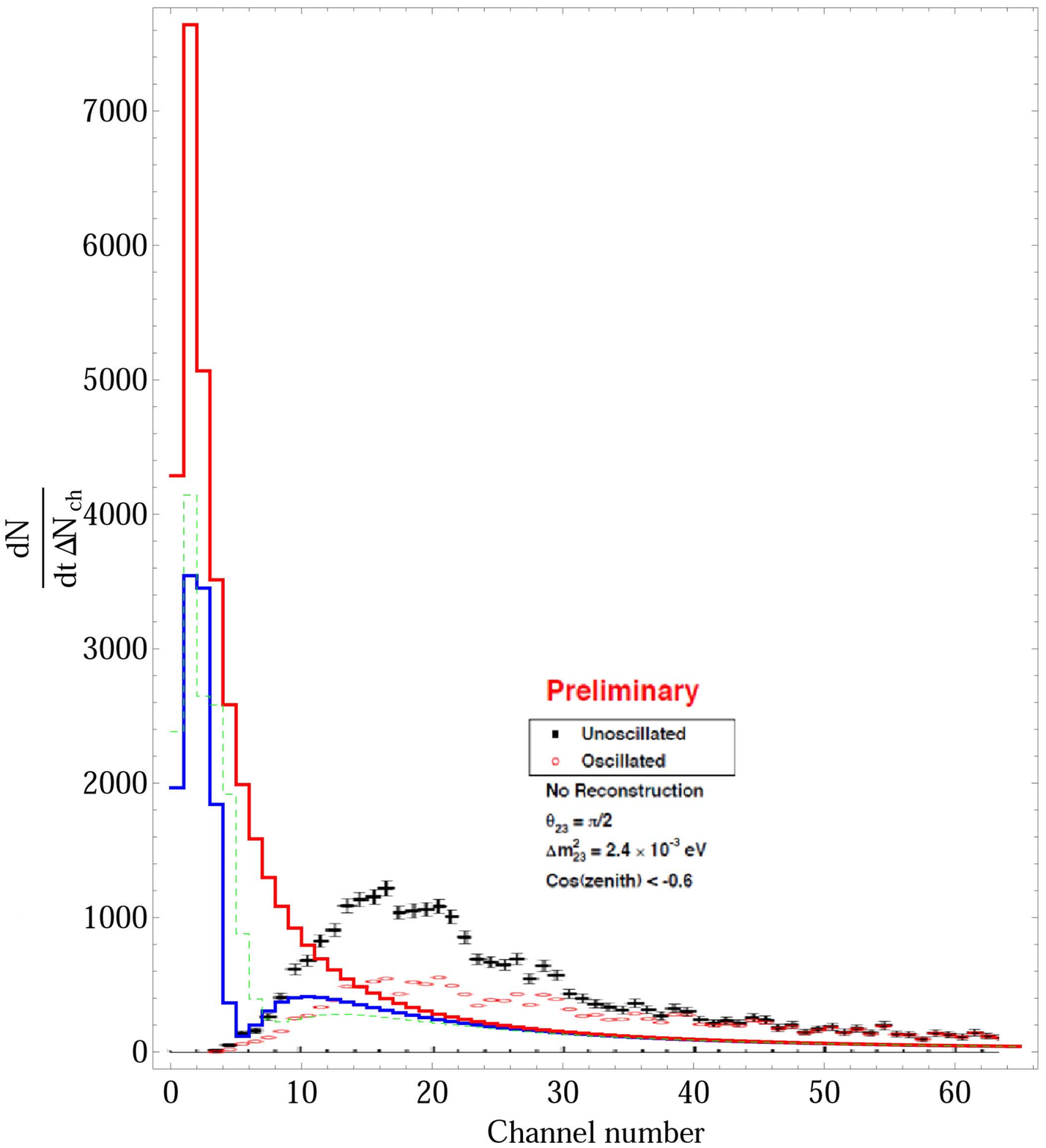}
\caption{\label{confronto}Our expected rate count rate (histograms) for upgoing muon tracks versus the most recent Deep Core expectations (dots) \cite{3}. As above the red curve marks the un--oscillated case, the blue one the CPT conserved version and the thin-dashed green line stand for the CPT violated solution. The discrepancy is relevant and we cannot find an explanation for it. Only above channel $50$ (where we cannot await for measures) the two rate maybe comparable; in all the channel range our estimate differ greatly in value and shape from the \cite{3}. The presence of additional neutral current shower events will increase channel 2-5 (almost doubling their number), but they maybe in principle disentangled from muon upward tracks by their uneven intensity ratio and their timing signature \cite{7}.}\label{confronto}
\end{center}
\end{figure}

We obtain thus the differential event number (per unit energy and solid angle and unit time) in function of neutrino energy and the horizontal  angle  $\theta$, in our definition starting from $0^{\circ}$ at horizon, equal to $90^{\circ}$ at vertical upward direction:

\begin{equation}
\frac{dN_{\mu}}{dE_{\nu}\,dt\,d\theta}(E_{\nu},\theta)\cdot\left(\frac{\Delta t}{\textrm{yr}}\right)\,=\,2\pi\,\frac{\phi_{\nu}(E_{\nu})}{E_{\nu}^2}\,f(E_{\nu},\theta)\,\sigma_{CC}(E_{\nu})\,N_A\,V(E_{\nu})\left(\frac{\Delta t}{\textrm{yr}}\right)\,P_{\nu_{\mu}\rightarrow\nu_{\mu}}(E_{\nu},\theta).
\end{equation}
\begin{figure}[h]
\begin{center}
\includegraphics[width=30pc]{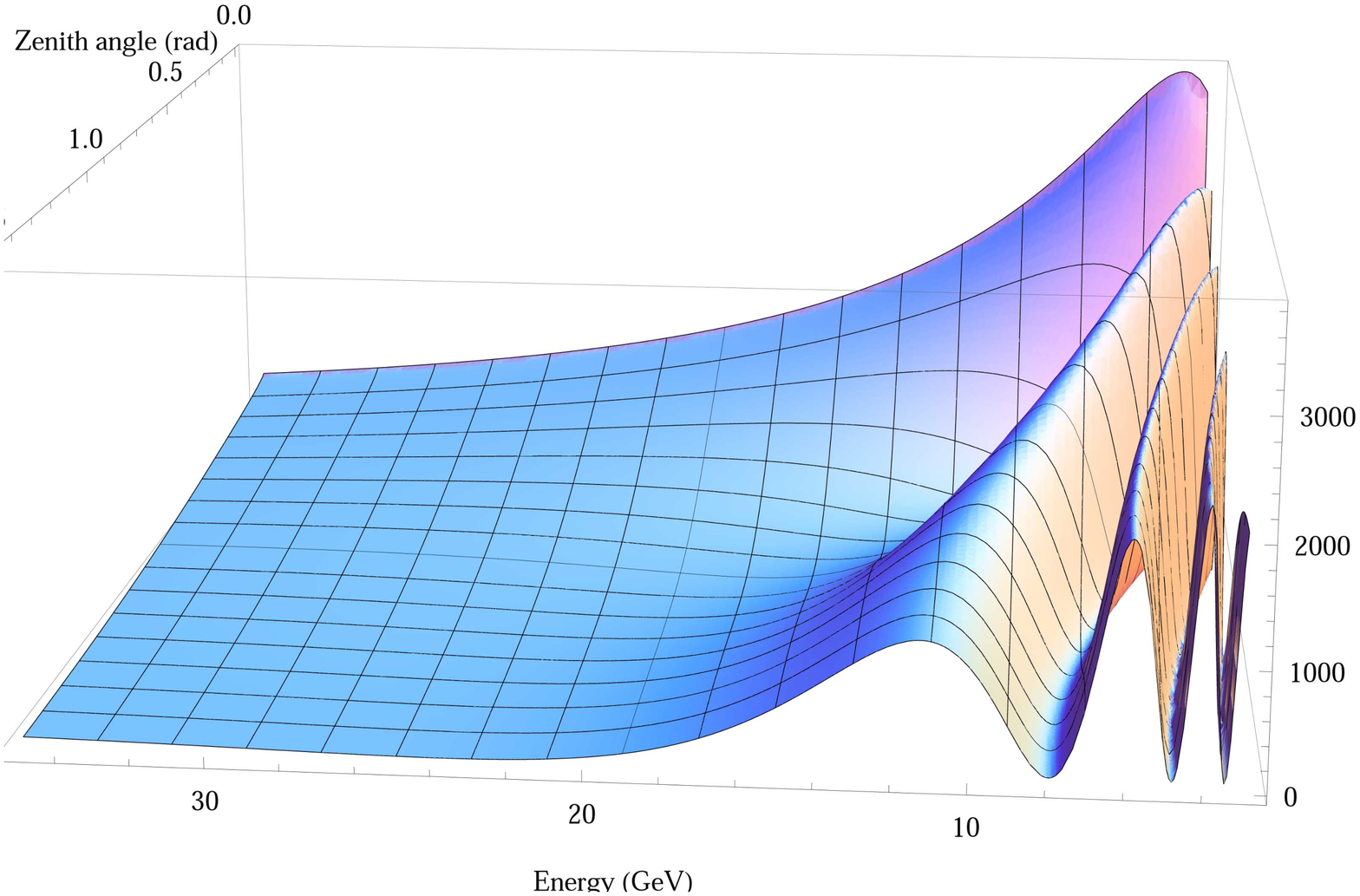}
\caption{\label{3D} The differential number event count a year in Deep Core as a function of the muon neutrino energy and angle, observable in narrow scale (top, $3$--$35$ GeV) assuming the vacuum among the neutrino flight).The horizontal axis is the energy in GeV, the vertical one is the event rate a year in Deep Core, the depth axis is the zenith angle variable in radiant.}
\includegraphics[width=32pc]{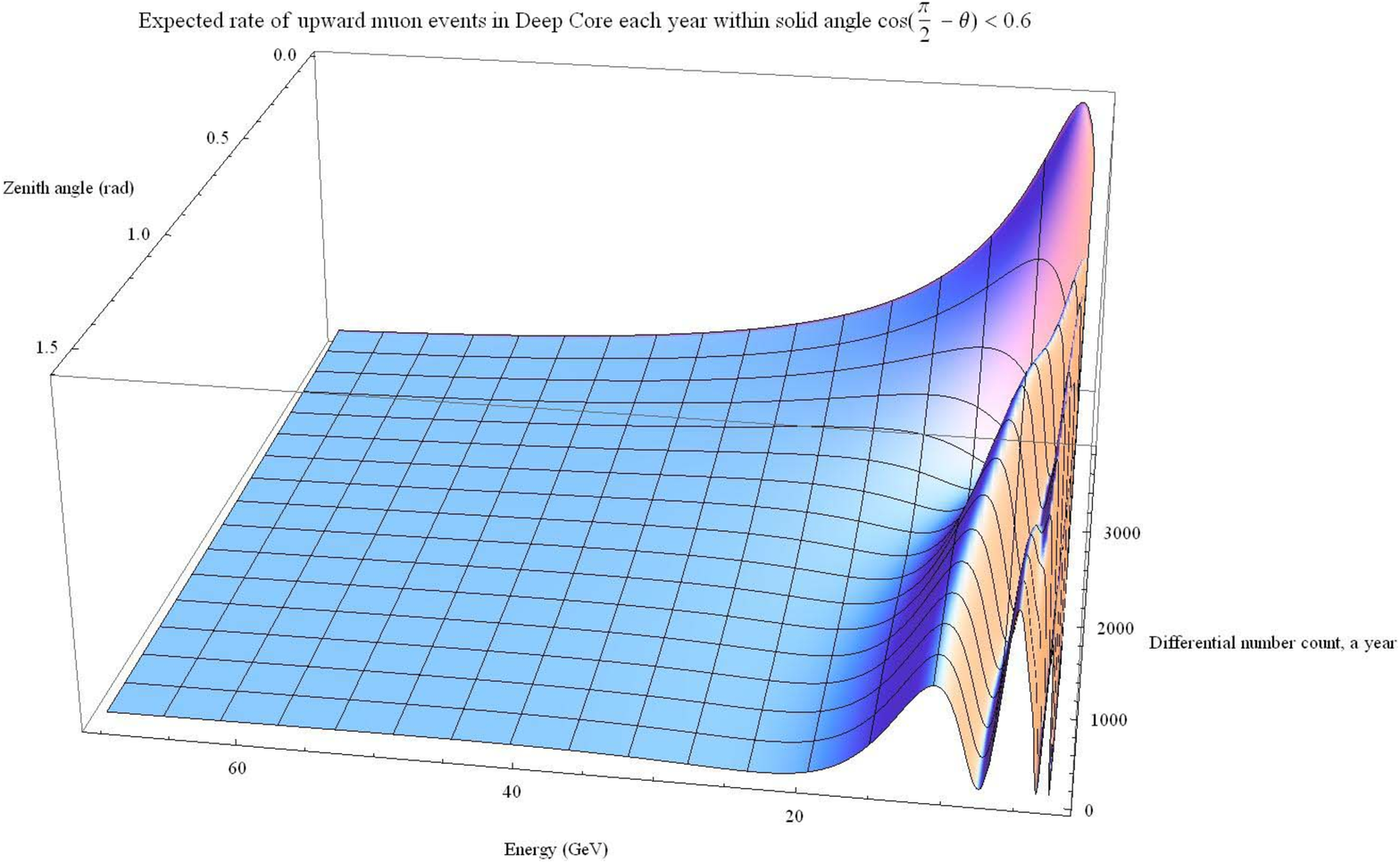}
\caption{\label{3D} As above the differential number event count a year in Deep Core as a function of the muon neutrino energy and angle, observable in wide scale (top, $3$--$65$ GeV) assuming an average Earth density $\rho = 7$ during up-going neutrino flight within Earth.The horizontal axis is the energy in GeV, the vertical one is the event rate a year in Deep Core, the depth axis is the zenith angle variable in radiant. Most of the inner oscillating structure is lost because of the Deep Core un-ability to reveal a few GeVs muon track. The last longest flux suppression at $20$ GeV survive and is source of a spectra deepening in Channel $6$--$9$}
\end{center}
\end{figure}
Event rate in one year of observation is obtained integrating over $\theta$ from $37^{\circ}$ below the horizon (equal to zenith angle $\cos(\hat{z})\,=\,\cos(\pi/2\,-\,\theta)\,=\,\sin(\theta)\,=\,0.6$ in Deep Core zenith angle) to $90^{\circ}$, and integrating over an appropriate energy bin:
\begin{equation}
\Delta N_{\mu}(E_{\nu})\,=\,\frac{dN_{\mu}}{dt}(E_{\nu})\,\left(\frac{\Delta t}{\textrm{yr}}\right)\,=\,\int_{E_i}^{E_i+\Delta E} \int^{\theta=90^\circ}_{\theta=37^\circ}\frac{dN_{\mu}}{dE_{\nu}\,dt\,d\theta}(E_{\nu},\theta)\,\cdot\,\left(\frac{\Delta t}{\textrm{yr}}\right)\,d\theta\,dE_{\nu},
\end{equation} where $N_A$ is the Avogadro number (multiplied by $10^6$, because ice density $\rho_{ICE}\,\approx\,10^3$ kg m$^{-3}$) and $V(E_{\nu})$ is the effective interaction volume in Deep Core (expressed in m$^3$, like eq. (\ref{V})).
Resulting events distribution are shown in figs. \ref{002GEV}, \ref{1GEV} and \ref{2e8GEV} for ideal narrow and real wide energy width. Later on the same spectra of muon tracks has to be projected along the vertical axis of Deep Core, as shown finally in fig. \ref{projected}.
The muon born in Charged Current interaction has nearly half the neutrino energy, and undergoes an average energy loss of $dE_\mu/dX\,\simeq\,200$ MeV m$^{-1}$ or $1/5$ GeV m$^{-1}$. Therefore the muon length (keeping in mind that kinematically holds $E_\mu\,\sim\,E_\nu/2$) in water is approximately:  $L_{\mu}\,=\,5\,E_{\nu}/2$ m, with energy as usual expressed in GeV. We used indeed the exact  linear-logarithmic growth in water:
\begin{equation}
L_{\mu}\,=\,3.9^{-1}\cdot10^4\,\ln\left(1\,+\,\frac{3.9}{4.0}\cdot10^{-3}\,E_{\nu_{\mu}}\right)\;\textrm{m}.
\end{equation}
To obtain a comparable plot of events distribution versus channel number we assumed an average projection factor, $\langle\cos(\pi/2\,-\,\theta) \rangle\,=\,\langle\sin\theta\rangle$ of muon track along the DOM strings due to the spread of the muon track direction (in range $0.6\,<\,\sin\theta\,<\,1$) first to be $0.8$ at same plane; however an additional projection factor $0.8$ is needed because muons tracks may be often skew respect the string line. The overall factor ranges  (in average between $0.8^2$ and $0.8$). We assumed the medium projection factor equal to $0.72$.
\begin{equation}
\Delta N_{\mu}(N_{ch})\,=\,\frac{dN_{\mu}}{dt}(E_{\nu})\,\left(\frac{\Delta t}{\textrm{yr}}\right)\,=\,\int_{E_i}^{E_i+\Delta E_{ch}/\langle\cos\theta\rangle}\int^{\theta=90^\circ}_{\theta=37^\circ}\frac{dN_{\mu}}{dE_{\nu}\,dt\,d\theta}(E_{\nu},\theta)\cdot\left(\frac{\Delta t}{\textrm{yr}}\right)\,d\theta\,dE_{\nu}.
\end{equation} Each channel to channel distance corresponds no longer to $2.8$ GeV (as for a vertical muon) but because of the inclined muon track projected along the vertical string $ <\sin{\theta}> \simeq 0.72$, on average, to a wider energy band of $2.8/0.72\,=\,3.88$ GeV.
This projecting factor smear the sharp decay on rate event due to the oscillation and it mask most of the low energy neutrino oscillation, whose inner traces has been revealed by SK  only in last decade. However the most severe deepening in the spectra (see our result versus earlier one in fig. \ref{projected}) at about $20$ GeV survive the smearing and it marks our expected count rate.

\section{Conclusions}
The Deep Core telescope rate should be soon published: their spectra (projected and counted along the channel) might be foreseen now and compared soon with the records.
Even if we did first used the Deep Core predictions \cite{6}, in late papers \cite{7} we did wonder and did suggest a different expected rate.
In the present paper we show step by step the procedure to foresee the Deep Core rate based on known Cosmic Ray rate and on SK muon neutrino flux,
flux interacting in Deep Core (variable) volume, leading, after mixing (see Appendix)  to survived muon (and anti-muons) neutrinos whose lepton tracks may rise in Deep Core Optical Module string. Our results differ from other ones \cite{4}, see fig \ref{confronto}; we tried to verify the estimates in different check (versus other experiments) and we believe that they are (at first approximation) correct.
The Deep Core rate shape, its intensity might be soon verified and tested. If the expected (CPT conserved estimate ) are met, Minos claim might be rejected. If at some channels (mostly 3-4-5) there will be an over production of signals, than the Minos CPT violated hint may be taken seriously.
It should be taken into account that NC (neutral current) events at tens GeV may spread light in spherical shape increasing and polluting the (3-6) channel rate; these a few thousands ($\simeq 3000-4000$) events may enrich channel 2-5; however their timing  structure  and their uneven  light intensity ratio may probe their different nature (see \cite{7}) from the up-going vertical muon ones, possibly to be separated easily.
Therefore these new Deep Core records may probe at minimum channel gate (6-8) an atmospheric suppression offering a rare windows for 20 GeVs neutrino astronomy. To observe such an astronomy in the whole arrival direction (not just along the zenith but also in its azimuth) we suggest the doubling of  the present Deep Core  array strings, allowing a contemporaneous  double string detection.  The discovery of twenty GeV neutrino astronomy in Deep Core may find a bonus also in rare, but nearby and powerful, Solar Flare neutrino brightening  \cite{10}

\section{Appendix: 3 flavor neutrino mixing}
The evolution equation for three neutrino mixing in matter reads :
\begin{equation}
i\frac{d}{d x}\Psi_{\alpha }=H \Psi_{\alpha } ,\ \  \text{with} \ \  H =\frac{1}{2 E}\left(U M U^{\dagger }+A\right),  \\\ \alpha =1,2,3
\end{equation}

which is a Schrodinger equation with a matter potential \(A\),

\[A=\left(
\begin{array}{ccc}
 A_{\text{CC}} & 0 & 0 \\
 0 & 0 & 0 \\
 0 & 0 & 0
\end{array}
\right),\text{ $\, \, \, $ }M=\left(
\begin{array}{ccc}
 0 & 0 & 0 \\
 0 & \text{$\Delta $m}_{21}^2 & 0 \\
 0 & 0 & \text{$\Delta $m}_{31}^2
\end{array}
\right),\text{ $\, \, \, $}\Psi _{\alpha }=\left(
\begin{array}{c}
 \psi _{\text{$\alpha $e}} \\
 \psi _{\alpha \mu } \\
 \psi _{\alpha \tau }
\end{array}
\right)\]

where \ $A_{CC}= 2 \sqrt{2}  E G_F N_e=0.76 \, \cdot 10^{-4} eV^2  ( \frac{E}{GeV}) (\frac{\rho}{g/cm^{3}}) $ is the potential only due to \(\nu _e\) charged current interaction.

This equation is a set of 9 differential equations whose solutions for probability conversion $P_{\alpha \beta}= \Psi_{\alpha \beta}^{2}$, with $\alpha, \beta =e,\mu, \tau$, has  boundaries conditions $P_{\alpha \beta}(0)$ equal to 1 for the starting flavor  $\mu$, 0 for the others (electron conversion into muon is negligible). This system of equations is symmetric since $P_{\alpha \beta}=P_{\beta \alpha}$.

 Among the simplest numerical solutions to show, we found a compact survival probability, for instance in vacuum for muon to muon, at a distance equal to the Earth diameter, corresponding to $\theta= 90^{\circ}$, as a function of neutrino energy: \\

 $P_{\mu \mu}= \left| 0.240128 + 0.304714\ e^{-2.5859 i/En} +
  0.455158\ e^{-75.9607 i/En} \right|^2 $

The more general cases taken into our study, with earth matter density included, are expression too wide and too complex (few pages of mathematica program) to be written here.

\section*{References}

\smallskip


\end{document}